\documentstyle[prl,aps,twocolumn]{revtex}

\begin{document}

\newcommand\ba{\begin{array}}
\newcommand\ea{\end{array}}
\newcommand\bc{\begin{center}}
\newcommand\ec{\end{center}}
\newcommand\be{\begin{enumerate}}  
\newcommand\ee{\end{enumerate}}  
\newcommand\bi{\begin{itemize}}  
\newcommand\ei{\end{itemize}}  
\newcommand\bd{\begin{description}}  
\newcommand\ed{\end{description}}  
\newcommand\beq{\begin{equation}}  
\newcommand\eeq{\end{equation}}  
\newcommand\beqa{\begin{eqnarray*}}  
\newcommand\eeqa{\end{eqnarray*}}  

\newcommand\gap{\smallskip}
\newcommand\eqbox[1]{\fbox{\rule[-.8em]{0em}{2.3em}$\displaystyle\ #1\ $}}

\newcommand{\dum}{\thechapter.}
\newcommand{\eq}[1]{Eq.\ (\ref{#1})}
\newcommand{\eqs}[2]{(\ref{#1}--\ref{#2})}
\newcommand\cf{{\em cf\ }}
\newcommand\eg{{\em e.g.,\ }}
\newcommand\etc{{\em etc}}
\newcommand\ie{{\em i.e.,\ }}
\newcommand\qed{\hfill {\bbox{ QED}}}

\newcommand\A{\op A}
\newcommand\B{\op B}
\newcommand\mathC{\mkern1mu\raise2.2pt\hbox{$\scriptscriptstyle|$}
                {\mkern-7mu\rm C}}
\newcommand\D{\op D}
\renewcommand\H{{\cal H}}                       
\renewcommand\P{\op P}                          
\newcommand\Q{\op Q}
\newcommand\U{\op U}
\newcommand\Z{{\rm \bf Z}}

\newcommand\abs[1]{\vert {#1}\vert}
\newcommand\av[1]{\langle#1\rangle}
\newcommand\bra[1]{\langle#1\vert\,}
\newcommand\bracket[2]{\langle#1\vert#2\rangle}
\newcommand\dby[1]{{d\over d#1}}
\def\ddde{\de^{(3)} }
\newcommand\down{\vert\!\downarrow\,\rangle}

\newcommand\half{{\frac 12}}
\newcommand\hhalf{{\textstyle\half}}
\newcommand\ioh{{i\over\hbar}}

\newcommand\ket[1]{\,\vert#1\rangle}
\newcommand\la{\langle}
\newcommand\lleft{\vert\!\leftarrow\rangle}
\newcommand\map{\longrightarrow}
\newcommand\mathR{{\rm I\! R}}
\newcommand\norm[1]{\parallel\!\v#1\!\parallel}
\newcommand\op[1]{\widehat{#1}}
\newcommand\ra{\rangle}
\newcommand\rright{\vert\!\rightarrow\rangle}
\newcommand\tr{{\rm tr}\,}
\newcommand\triple[3]{\langle#1\vert\,#2\,\vert#3\rangle}
\newcommand\twid[1]{\tilde {#1}}
\newcommand\unit{{\rm I}}
\newcommand\up{\vert\!\uparrow\,\rangle}
\renewcommand\v[1]{{\bf{#1}}}        

\newcommand\<[2]{\langle\v#1,\v#2}              
\renewcommand\>{\rangle}                        
\renewcommand\[{[\,}                            
\renewcommand\]{\,]}                            

\renewcommand\a{\alpha}                         
\renewcommand\b{\beta}                          
\newcommand\g{\gamma}
\newcommand\de{\delta}
\newcommand\e{\varepsilon}
\newcommand\z{\zeta}
\newcommand\k{\kappa}
\renewcommand\l{\lambda}                        
\newcommand\m{\mu}
\newcommand\n{\nu}
\newcommand\r{\rho}
\newcommand\s{\sigma}
\newcommand\th{\theta}
\newcommand\f{\phi}
\newcommand\w{\omega}
\renewcommand\O{\Omega}				
\renewcommand\L{\Lambda}			
\newcommand\bolde{\epsilon\mkern-6mu\epsilon}  
\newcommand\nnabla{\nabla\mkern-14mu\nabla}  
\newcommand{\mue}{{\mbox{\boldmath $\mu$}}}

\def\pdby#1{\partial\over\partial #1}
\def\exl{\raise1pt\hbox{$\scriptstyle<$}}
\def\exr{\raise1pt\hbox{$\,\scriptstyle>$}}

\newcommand\rem[1]{$\vert \underline{\overline{\text{{\bf#1}}}}\vert
$}
\newcommand\dmu{\partial_\mu}

\draft

\preprint{{\tt Imperial/TP/97-98/26}} 

\title{A way to get a well-defined derivative expansion of real-time thermal effective
actions\thanks{{\tt talk presented at the 5th International Workshop on Thermal 
Field Theories and Their Applications, Regensburg, Germany, August 10-14, 
1998}}}

\author{M.Asprouli\thanks{email: {\tt m.asprouli@ic.ac.uk}}, V. Galan-Gonzalez\thanks{email:{\tt v.galan-gonz@ic.ac.uk}}}
\address{Theoretical Physics, Blackett Laboratory, Imperial College,
Prince Consort Road, London SW7 2BZ,  U.K. }
\date{14th September 1998}

\maketitle 

\begin{abstract}
We compute the quadratic part of the thermal effective action for real scalar 
fields which are initially in thermal equilibrium and vary slowly in time 
using a generalised real-time formalism proposed by Le Bellac and Mabilat 
\cite{belmab}. 
We derive both Real Time and Imaginary Time Formalisms and find that the 
result is analytic at the limits of zero external four-momenta when using 
our full time contour. We expand the fields in time up to the second 
derivative and discuss the initial time dependence of our result before 
and after the 
expansion in terms of equilibrium.

\end{abstract}


\section{Introduction}

We consider a two real scalar field theory with fields $\phi,\eta$ with 
Lagrangian ${{\cal L}}$ given by
\beq
\label{eqn1}
{\cal L} [\phi,\eta]=\frac 1{2}\partial_{\mu}\eta\partial^{\mu}\eta-\frac 1{2}m^2{\eta}^2-\frac 1{2}g\phi{\eta}^2+{{\cal L}_0}
\eeq
where ${{\cal L}_0}$ denotes the free Lagrangian for $\phi$.
If we integrate out the
$\eta$-field fluctuations and use a one-loop approximation we find
that the generating functional can be expressed as 
\beq
\label{eqn2}
{\cal Z}=\int_C{\cal D}\phi e^{iS_0 [\phi ]+iS_{eff} '[\phi ]}
\eeq
where $S_0$ is the classical action and $S_{eff} '$ is given by
\beq
\label{eqn3}
S_{eff} '[\phi ]=\frac i{2}\mbox{Trln}[1-gD_c\phi]
\eeq
In \eq{eqn3} $D_c$ is the propagator for the $\eta$ field and
\beq
\label{eqn4}
Tr=\int_{C} dt\int d^{3}\v{x}
\eeq
where the time path $C$ starts at initial time $t_i$ when the field $\eta$ 
is in equilibrium and ends at time $t_i-i\b$ where $\b$ is the inverse 
temperature. The path is the one of Fig.1 where $t_0$ is the time around 
which we will expand our time-dependent field. It consists in two parts, the 
horizontal $C_H$ $[t_i,t_f]$ and $[t_f,t_i]$ and the vertical $C_V$ from 
$[t_i,t_i-i\b]$.
\begin{figure}[htb]
\begin{center}
\setlength{\unitlength}{0.5pt}
\begin{picture}(495,280)(35,480)
\put( 70,565){\makebox(0,0)[lb]{\large $C_V$}}
\put(185,750){\makebox(0,0)[lb]{\large $\Im m(\tau)$}}
\put(250,710){\makebox(0,0)[lb]{\large $C_H$}}
\put(510,645){\makebox(0,0)[lb]{\large $\Re e (\tau)$}}
\put(510,700){\makebox(0,0)[lb]{\large $t_f$}}
\put(400,705){\makebox(0,0)[lb]{\large $t_0$}}
\put( 50,705){\makebox(0,0)[lb]{\large $t_i$}}
\put(70,490){\makebox(0,0)[lb]{\large $t_i-i \beta$}}
\thicklines
\put(400,690){\circle*{10}}
\put( 40,680){\vector( 1, 0){490}}
 
\put( 60,690){\vector( 1, 0){220}}
\put(280,690){\line( 1, 0){220}}
\put(500,680){\oval(10,20)[r]}
\put(500,670){\vector(-1, 0){220}}
\put(280,670){\line(-1, 0){220}}
\put( 60,670){\vector( 0,-1){110}}
\put( 60,560){\line( 0,-1){ 60}}
\put(180,480){\vector( 0, 1){280}}
\end{picture}
\end{center}
\caption{The integration contour C in the complex t-plane.}
\end{figure}
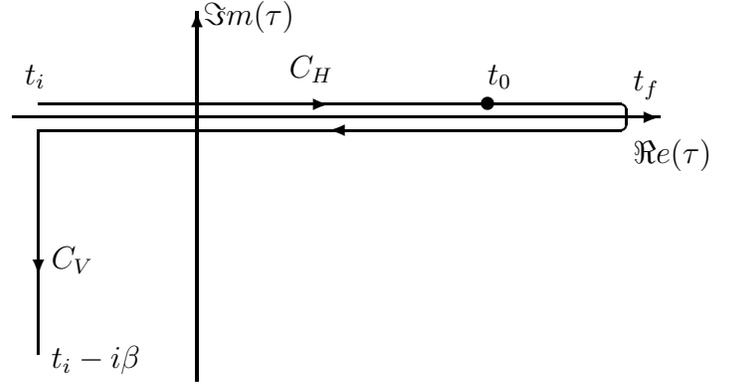
We expand the logarithm in $S_{eff} '$ and look at the first non-local 
term which is quadratic in the fields $\phi$. If we separate the time 
integration which interests us at finite temperature and Fourier transform 
in space, the quadratic term $S_{eff}^{(2)}$ is given by
\begin{eqnarray}
\label{eqn5}
\lefteqn{S_{eff}^{(2)}=
-\frac{ig^2}{4}\int_{t_i}^{t_i-i\beta}dt_0\int\frac{d^{3}\v{p}}
{{(2\pi)}^3}\int\frac{d^3\v{k}}{{(2\pi)}^3}}\nonumber\\
& &\times\int_{t_i}^{t_i-i\beta}dt_1
\,{\phi (\v{p},t_0) 
D_c (\v{k};t_0,t_1)\phi (\v{p},t_1)
D_c(\v{k}+\v{p};t_1,t_0)}\nonumber\\
& &
\end{eqnarray}    
We will refer to the $t_1$-integral as the ``bubble term''. This has been 
found to be non-analytic at the zero limits of the four external momenta 
$(p_0\rightarrow 0, {\bf p}\rightarrow 0)$ at finite temperature using 
the Real Time Formalism as shown by Das \cite{das} and others. It is 
associated with problematic delta functions with the same argument. It also 
occurs in hot gauge theories \cite{grib,sili,weld2} and has been related to 
the difference between the electric screening mass of the photon and the 
frequency of the plasma oscillations. In this spirit the expansion of the 
effective action in powers of momentum around the non-analytic origin and 
the lowest order in this expansion which is the effective potential would be 
not well-defined.

In the following we will deal with these problems using our generalised time
contour and we will recover RTF and ITF in the appropriate limits of 
the initial time $t_i$. We will expand our field in time up to the 
second order and we will compare our result with recent calculations by Evans 
\cite{eva3} and Berera et al \cite{ber}. We will discuss the
$t_i$-dependence 
of the ``bubble'' term before and after the expansion in terms of equilibrium
and the periodicity of the fields.

\section{The method}
The propagators used are in their mixed representation $D_c(t,\v{k})$ by 
Mills \cite{mill} 
\beq
\label{eqn6}
D_c(t,\v{k})=\int \frac{d k_0}{2\pi}e^{-i k_0
t}\[\th_c(t)+n(k_0)\]\rho(k_0,\v{k})
\eeq
where $\theta_{c}(t)$ is a contour $\theta$ function, $n(k_{0})$ is the 
Bose-Einstein distribution function given by
\beq
\label{eqn7}
n(k_0)=\frac1{e^{\b k_0}-1}
\eeq
 and $\rho(k_{0},{\bf{k}})$ is the 
(temperature independent) two-point spectral function given by
\beq
\label{eqn8}
\rho(k_0,\v{k})=2\pi\e(k_0)\de(k_0^2-\w_k^2)
\eeq
where $\e(k_0)$ is the sign function and
\beq
\label{eqn9}
\w_k^2=\v{k}^2+m^2
\eeq
The time contour $C$ is the one of Fig.1 where the horizontal path 
$C_H$ consists in the parts $[t_i,t_0]$, $[t_0,t_f]$, $[t_f,t_0]$ and
$[t_0,t_i]$ and the vertical path $C_V$ is the $[t_i,t_i-i\b]$ part. 
The ITF is recovered when $t_i=t_0$ where only the vertical path survives 
and the RTF when $t_i\rightarrow -\infty$ where only the horizontal path 
survives. 
In \eq{eqn6} we write the delta functions of $\rho(k_0,\v{k})$ in their 
regularised form in order to be able to take the RTF limit and thus 
$\rho(k_0,\v{k})$ is written as
\beq
\label{eqn10}
\rho(k_0,\v{k})=\frac i{2\w_k}\sum_{r,\xi=\pm1}\frac{r\xi}{k_0-\xi\w_k+ir\e }
\eeq
where $\e$ is the regulator which we keep finite until the 
$t_i\rightarrow -\infty$ limit has been performed.
We Taylor expand the field $\phi (\v{p},t_1)$ around $\phi (\v{p},t_0)$ and 
write it as an exponential of the time derivative $\partial_t$ acting on 
$\phi (\v{p},t)$
\begin{eqnarray}
\label{eqn11}
\phi(\v{p},t_1)&=&\sum_{n=0}^\infty\left. \frac 1{n!}(t_1-t_0)^n\frac{\partial^n}{\partial_t^n}\phi(\v{p},t)\right\vert_{t=t_0}\!\ \nonumber\\
&=&\left. e^{(t_1-t_0)\partial_t}\phi(\v{p},t)\right\vert_{t=t_0}
\end{eqnarray}
Substituting in \eq{eqn5} the ``bubble'' term is written as
\begin{eqnarray}
\label{eqn12}
\lefteqn{\Gamma^{(B)}=}\nonumber\\
& &\int_C dt_1\phi(\v{p},t_0)D_c(t_0,t_1)\left. e^{(t_1-t_0)\partial_t}\phi(\v{p},t)\right\vert_{t=t_0}  D_c(t_1,t_0)\nonumber\\
& &
\end{eqnarray}
If we expand the exponential in powers of the time derivative $\partial_t$, 
$\Gamma^{(B)}$ is also written as 
\begin{eqnarray}
\label{eqn13}
\lefteqn{\Gamma^{(B)}=}\nonumber\\
& &\int_{t_i}^{t_i-i\beta}dt_1\phi(\v{p},t_0)\left. 
\,[\Gamma^{(0)}+\Gamma^{(1)}\partial_t+\Gamma^{(2)}\partial_t^2+...]
\,\phi(\v{p},t)\right\vert_{t=t_0}\nonumber\\
& &
\end{eqnarray}
In the following we will find the zeroth order term $\Gamma^{(0)}$ in the 
$\partial_t$ expansion which is related to the effective potential 
and the coefficients 
$\Gamma^{(n)}$ of the higher order derivatives up to the second order. 

In \eq{eqn12} the exponential $e^{(t_1-t_0)\partial_t}$ acts as an energy 
shift by $-i\partial_t=p^0$ in the energies of the propagators where 
$p^0$ is the external energy. The energies $\Omega,\w$ in the dispersion 
relations of the propagators are related to the three momenta $ \v{k}+\v{p}, 
\v{k}$ respectively where $\v{p}=-i\nnabla$ is the external three momenta. 
Thus the zero four 
momenta limits correspond to taking the limits $\partial_t\rightarrow 0
\,(p^0\rightarrow 0)$ and $\Omega\rightarrow \pm\w\,(\v{p}\rightarrow 0)$.

\section{The bubble term}
We perform the $t_1$ integration in \eq{eqn12} along the whole time path for 
${\it both}$ horizontal and vertical parts and get 
\begin{eqnarray}
\label{eqn14}
\lefteqn{\Gamma^{(B)} =\sum_{\pm\w,\Omega}\frac{in(\w)n(\Omega)}{4\w\Omega}
\times}\nonumber\\
& &\times[\frac{(e^{\beta(\w+\Omega)}-1)}{A}-e^{-iA\Delta t}\frac
{(e^{-i\b\partial_t}-1)}{A}]
\end{eqnarray}
where $$A=\w+\Omega+i\b\partial_t$$ and $$\Delta t=t_i-t_0$$
This result agrees with the one found by Evans using ITF \cite{eva3}. 
We notice that the $t_i$-dependence is included in the second term of 
\eq{eqn14} which is being multiplied by $(e^{-i\b\partial_t}-1)$. But 
$$e^{-i\b\partial_t}\phi(t)=\phi(t-i\b)$$ and if the field $\phi(t)$ is 
periodic it is also $$\phi(t-i\b)=\phi(t)$$ This means that 
$$(e^{-i\b\partial_t}-1)\phi(t)=0$$ and the $t_i$-dependent term vanishes 
provided that we have included both horizontal and vertical paths of the 
contour. This agrees with Le Bellac and Mabilat \cite{belmab} who showed the 
$t_i$-independence for the case of the effective potential.

\section{Analyticity}
We perform the zero limits of the external four momenta 
$(\Omega\rightarrow \pm\w\,,\partial_t\rightarrow 0)$ 
in both orders in the result of \eq{eqn14}. In the case of $\Delta t$ finite, 
when both paths of the time contour contribute, the result is analytic and 
is the usual effective potential \cite{das} given by
\begin{eqnarray}
\label{eqn15} 
\Gamma^{(0)}=i\frac{(2n(\w)+1)}{2\w^3}+i\frac{\beta n(\w)(1+n(\w))}{w^2}
\end{eqnarray}
This is also the result for the ITF case of $\Delta t=0$. For 
$\Delta t\rightarrow -\infty$ which is the RTF limit, we recover the result 
of \eq{eqn15} ${\it only}$ when performing the time-derivative limit 
$\partial_t\rightarrow 0$ first while in the opposite order of limits we get 
$$
\lim_{\Omega\rightarrow \pm\w,\partial_t\rightarrow 0}\Gamma ^{(B)}= 
i\frac{(2n(\w)+1)}{2\w^3} $$ which is only the first term of the full
result of \eq{eqn15}, recovering thus the non-analyticity using
RTF. It is therefore essential that we use both paths of the time
contour to get a well-defined analytic result.

\section{Time-derivative expansion}
For both $\Delta t$ cases we first take the limits $\Omega\rightarrow \pm\w$ in
\eq{eqn14} and then expand in powers of the time-derivative $\partial_t$ 
to get the coefficients $ \Gamma^{(n)}$ up to the second order.
\be
\item $\Delta t\rightarrow -\infty$ case

Performing the first limit ($\Omega\rightarrow \pm\w$), \eq{eqn14} gives
$$
\lim_{\Omega\rightarrow \pm\w}\Gamma ^{(B)}=2i\frac{(2n(\w)+1)}{\w(4\w^2+\partial_t ^2)}
$$
and after expanding in powers of the time-derivative $\partial_t$, we get a 
zeroth order in the time-derivative term which is the first part of the 
effective potential in 
agreement with before and a second order term of the form 
$$
\Gamma^{(2)}(\Delta t\rightarrow -\infty)= -i\frac{(2n(\w)+1)}{8\w^5}
$$
which agrees with a recent result by Berera et al \cite{ber}. In this case we did not 
find any coefficient $ \Gamma^{(1)}$ of the term proportional to the single 
time-derivative of the field.
\item $\Delta t$ finite case

The expansion in powers of $\partial_t$ gives us a zeroth term which is the 
full effective potential of \eq{eqn15} as shown before, and a second order 
term which, in the limit of $\Delta t=0$, is
\beqa
\lefteqn{\Gamma^{(2)}(\Delta t=0)=}\\
& & -\frac i{8\w^5}\[(2n(\w)+1)-\beta\w(2n(\w)(n(\w)+1)+1)\\
& & \mbox{}+\b^2\w^2(2n(\w)+1)+\frac{4\b^3\w^3n(\w)(n(\w)+1)}{3}]=\\
& & \Gamma^{(2)}(\Delta t\rightarrow -\infty)+{\it other\,terms}
\eeqa
This result is consistent with the one by Evans using ITF \cite{eva3} 
but it is different to the one for $\Delta t$ infinite proving the 
$t_i$-dependence of the second order term $\Gamma^{(2)}$.
 
But in the finite $\Delta t$ case we ${\it also}$ had a 
coefficient $\Gamma^{(1)}$ for the 
single time-derivative of the field unlike the 
zero temperature case where such a term vanishes. Evans \cite{eva3} found 
a similar linear term in his imaginary time calculations. 
In the case of $\Delta t=0$ (ITF) it is related to the effective 
potential $\Gamma^{(0)}$ by
$$\Gamma ^{(1)}=-\frac{i\beta}{2}\Gamma ^{(0)}$$
The existence of this term is due to the extra four-velocity with respect 
to the heat bath at finite temperature. The invariant quantity now is 
$U_{\mu}\partial^{\mu}$ where $$U_{\mu}=(1,0,0,0)$$ is the rest frame of 
the heat bath. Such a term did not exist in the expansion for the 
$\Delta t$ infinite case,
 where only zero and second order terms in the time-derivative
survived. This makes sense since in the infinite time limit any interaction with the 
heat bath which gives rise to such linear terms will have been damped.
 Mathematically this term could arise due to the shape of the time contour,
 which in the finite
$\Delta t$ case is non-symmetric. However this is not the case for the 
zero-temperature situation or the finite temperature one in the infinite 
$\Delta t$ limit where the symmetry of the contour will make any time 
integration of odd terms in the derivative expansion to vanish. 
This term is also $t_i$-dependent. We know that the effective action 
should be independent of the initial 
time for 
fields in thermal equilibrium. However in our case we have expanded 
time-dependent fields which have almost departed from equilibrium in order for 
such an expansion to be meaningful and therefore a dependence on the initial 
time is expected. Moreover a truncated expansion even of periodic fields is 
not necessarily periodic.   
\ee
\section{Conclusions and future work}
We calculated the quadratic part of the thermal effective action for real 
scalar fields which vary slowly in time. We found the result to be analytic 
in the limits of zero external four-momenta as long as we consider both 
horizontal and vertical paths of our time contour ($\Delta t$ finite). We 
recovered the non-analyticity occuring in RTF as well as the ITF result in the 
appropriate limits of $t_i$. We also found a simple way of computing higher 
derivative 
terms in the bubble and derived the complete bubble term. 
We expanded the bubble term up to the second derivative in the fields and 
our results were consistent with recent RTF and ITF calculations. 
The non zero, linear time-derivative term found in the 
finite $\Delta t$ case is related to the heat bath 
frame at finite temperature. The physical meaning of
such a term and in particular its sign and whether it is complex or real 
 will determine whether it should be considered as a dissipative term 
or as a kind of ``chemical potential''. We can solve the equations of 
motion for the $\phi$ field in 
the truncated expansion and study the importance of this term for the dynamics 
of the field.

We also studied the $t_i$-dependence of our result before and after the 
expansion and discussed it in terms of equilibrium and the truncated expansion 
considered. We found that  ${\it only}$ for periodic $\phi$ fields does the 
$t_i$-dependence cancel provided that we consider both horizontal and
vertical paths of the time contour. 

The extension of our calculation to higher derivative terms and to 
space-dependent fields will give the full effective action but the spatial 
contribution is quite trivial since the problems arising at finite temperature 
involve time. 
We have considered a two real scalar field theory, but we have also used 
our method in more physical models, such as a Yukawa theory, where the fermion 
field is the one integrated out and we have found a similar ``linear term''. 
We can also consider scalar QED and gauge theories or even consider systems 
with time-dependent parameters. The 
possibility of evaluating quantum corrections can be 
more directly applied to phase transitions, where they
may indicate us something about the order of the transition.


\end{document}